\documentclass[prl,noshowpacs,english,superscriptaddress,twocolumn]{revtex4}
\usepackage{amsmath,amssymb,amsfonts,mathrsfs}
\usepackage{amsthm}
\usepackage{CJK}
\usepackage{bm}
\usepackage{babel}
\usepackage{graphicx}
\allowdisplaybreaks
\usepackage{braket}
\usepackage[breaklinks]{hyperref}
\usepackage{color}
\hypersetup{colorlinks=true, linkcolor=blue, citecolor=blue, filecolor=blue, urlcolor=blue}

\begin{document}

\title{Two-dimensional Dirac Semimetals without Inversion Symmetry}

\author{Y. J. Jin}
\affiliation{Department of Physics $\&$ Institute for Quantum Science and Engineering, Southern University of Science and Technology, Shenzhen 518055, P. R. China.}
\affiliation{Department of Physics, The University of Hong Kong, Pokfulam Road, Hong Kong}

\author{B. B. Zheng}
\affiliation{College of Physics and Optoelectronic Technology $\&$ Advanced Titanium Alloys and Functional Coatings Cooperative Innovation Center, Baoji University of Arts and Sciences, Baoji, 721016, P. R. China}

\author{X. L. Xiao}
\affiliation{Department of Physics $\&$ Institute for Quantum Science and Engineering, Southern University of Science and Technology,
Shenzhen 518055, P. R. China.}
\affiliation{Department of Physics, Chongqing University, Chongqing 400044, P. R. China.}

\author{Z. J. Chen}
\affiliation{Department of Physics $\&$ Institute for Quantum Science and Engineering, Southern University of Science and Technology,
Shenzhen 518055, P. R. China.}
\affiliation{Department of Physics, South China University of Technology, Guangzhou 510640, P. R. China}

\author{Y. Xu}
\affiliation{Department of Physics $\&$ Institute for Quantum Science and Engineering, Southern University of Science and Technology,
Shenzhen 518055, P. R. China.}

\author{H. Xu}
\email[]{xuh@sustech.edu.cn}
\affiliation{Department of Physics $\&$ Institute for Quantum Science and Engineering, Southern University of Science and Technology,
Shenzhen 518055, P. R. China.}

\begin{abstract}
Realizing stable two-dimensional (2D) Dirac points against spin-orbit coupling (SOC) has attracted much attention because it provides a platform to study the unique transport properties. In previous work, Young and Kane [Phys. Rev. Lett. \textbf{115}, 126803 (2015)] proposed stable 2D Dirac points with SOC, in which the Berry curvature and edge states vanish due to the coexistence of inversion and time-reversal symmetries. Herein, using the tight-binding model and k$\cdot$p effective Hamiltonian, we present that 2D Dirac points can survive in the presence of SOC without inversion symmetry. Such 2D Dirac semimetals possess nonzero Berry curvature near the crossing nodes, and two edge states are terminated at one pair of Dirac points. In addition, according to symmetry arguments and high-throughput first-principles calculations, we identify a family of  ideal 2D Dirac semimetals, which has nonzero Berry curvature in the vicinity of Dirac points and visible edge states, thus facilitating the experimental observations. Our work shows that 2D Dirac points can emerge without inversion symmetry, which not only enriches the classification of 2D topological semimetals but also provides a promising avenue to observe exotic transport phenomena beyond graphene, e.g., nonlinear Hall effect.

\end{abstract}

\pacs{73.20.At, 71.55.Ak, 74.43.-f}

\keywords{ }

\maketitle

\begin{figure}
	\centering
	\includegraphics[scale=0.042]{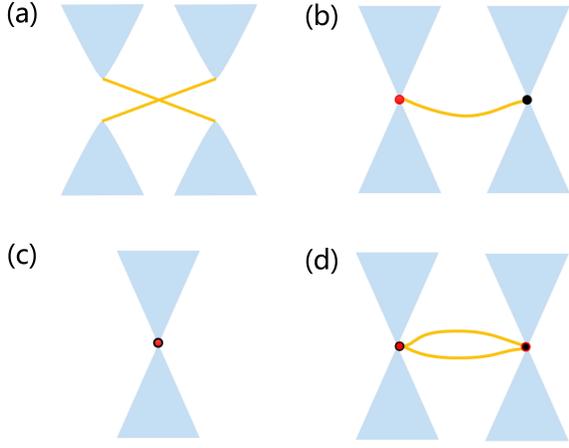}
	\caption{The classification of crossing points in two dimensions with SOC and $\mathcal{T}$ symmetry. Light yellow lines denote the corresponding edge states. Red and black dots represent the crossing points with opposite chirality. (a) Dirac points in graphene, which are destroyed by SOC. (b) Non-symmorphic symmetry protected Weyl points against SOC. (c) $\mathcal{PT}$ symmetry-involved Dirac points without edge states. (d) Dirac points without $\mathcal{P}$ symmetry characterized by two edge states.
\label{figure-1}}
\end{figure}

Dirac semimetals with fourfold degenerate points (i.e., Dirac nodes) near the Fermi level possess intriguing physical properties, such as unique Fermi arcs, ultrahigh mobility, giant magnetoresistance and Klein tunelling\cite{RevModPhys.90.015001,Cd3As2,Klein}. As the beginning of Dirac materials, graphene has attracted significant attention, partially due to its Dirac cones at the Fermi level and associated electronic properties\cite{graphene1,graphene2}. In the absence of spin-orbit coupling (SOC), such massless Dirac cones are protected by inversion ($\mathcal{P}$) and time-reversal ($\mathcal{T}$) symmetries, where the spin rotation symmetry is conserved. However, once the SOC is included, the spin rotation symmetry is broken that drives graphene into a quantum spin Hall insulator (QSHI) with a pair of helical edge states inside the energy gap\cite{grapheneTI1,grapheneTI2}, as shown in Fig. \ref{figure-1}(a). Benefitting from the weak SOC, graphene is regarded as a two-dimensional (2D) Dirac semimetal, while other candidate materials\cite{Ge1,Sn1,ncorg} that are composed of heavier elements turn into QSHIs with visible energy gaps. As a result, the study of 2D Dirac semimetals is limited to graphene. To further explore the unique properties of 2D Dirac semimetals, it is essential to investigate robust Dirac points against SOC.

Physically, non-symmorphic symmetries can protect the band crossings along high-symmetry lines or at high-symmetry points in electronic band structures\cite{bradley1976p,Nature2hourglass,ben1,ben2,ben3,kru}. In the presence of $\mathcal{T}$ symmetry and SOC, one non-symmorphic symmetry in two dimensions, such as screw axis, glide mirror line, or glide mirror plane, can lead to the formation of 2D Weyl nodes\cite{np2Dcrossing,Kane2dDirac} because non-symmorphic operators have higher-dimensional projective representations. Analogous to three-dimensional (3D) case, such 2D Weyl nodes are characterized by the local Chern number or quantized Berry phase and finite edge states terminated at the projections of two Weyl points in one-dimensional (1D) Brillouin zone (BZ)\cite{Kane2dDirac,Diracprb,prbgs}, as shown in Fig. \ref{figure-1}(b). When the $\mathcal{P}$ symmetry is further introduced, the $\mathcal{PT}$ symmetry enforces one pair of Weyl points with opposite chirality to merge at the time-reversal invariant momentum (TRIM) points, forming robust Dirac points against SOC\cite{Kane2dDirac}. However, in such 2D Dirac semimetals, the Berry curvature must disappear throughout the whole BZ under the $\mathcal{PT}$ symmetry\cite{Kane2dDirac,Berryphase}. Meanwhile, the edge states also vanish, as shown in Fig. \ref{figure-1}(c), and they can appear again if the Dirac point is split into one pair of Weyl nodes by breaking either $\mathcal{P}$ or $\mathcal{T}$ symmetry. It is well known that Berry curvature is closely related to many physical phenomena, and nontrivial edge states exhibit unique electronic transport properties\cite{TI1,TI2,transportgraphene,wanweyl,weyltransport,weyltransport1}. Hence, it is of particular importance to construct stable 2D Dirac points without $\mathcal{P}$ symmetry.

It is generally believed that the crossing of two Kramers degenerate bands forms the fourfold degenerate Dirac point. In other words, the $\mathcal{P}$ and $\mathcal{T}$ symmetries should be preserved to obtain the stable Dirac point. However, as discussed above, to realize stable Dirac points against SOC with nonzero local Berry curvature, it is necessary to break $\mathcal{P}$ or $\mathcal{T}$ symmetry. Once $\mathcal{P}$ or $\mathcal{T}$ symmetry is broken, other symmetries should be introduced to guarantee the fourfold degenerate band crossing. Therefore, it is natural to recall the fact that non-symmorphic symmetries can bring extra degeneracies and therefore give rise to symmetry-protected band crossings in two dimensions.

Here, we theoretically propose robust 2D Dirac points against SOC in layer group \emph{Pb2b} (No. 30), which are protected by the combination of $\mathcal{T}$ and glide mirror symmetries. Different from $\mathcal{PT}$ symmetry-involved 2D Dirac points\cite{Kane2dDirac}, the 2D Dirac points without $\mathcal{P}$ symmetry are characterized by nonzero Berry curvature in the vicinity of Dirac points, and two edge states connect one pair of Dirac points [see Fig. \ref{figure-1}(d)]. Furthermore, a family of ideal candidates is identified to realize such robust 2D Dirac points.

\begin{figure}
	\centering
	\includegraphics[scale=0.12]{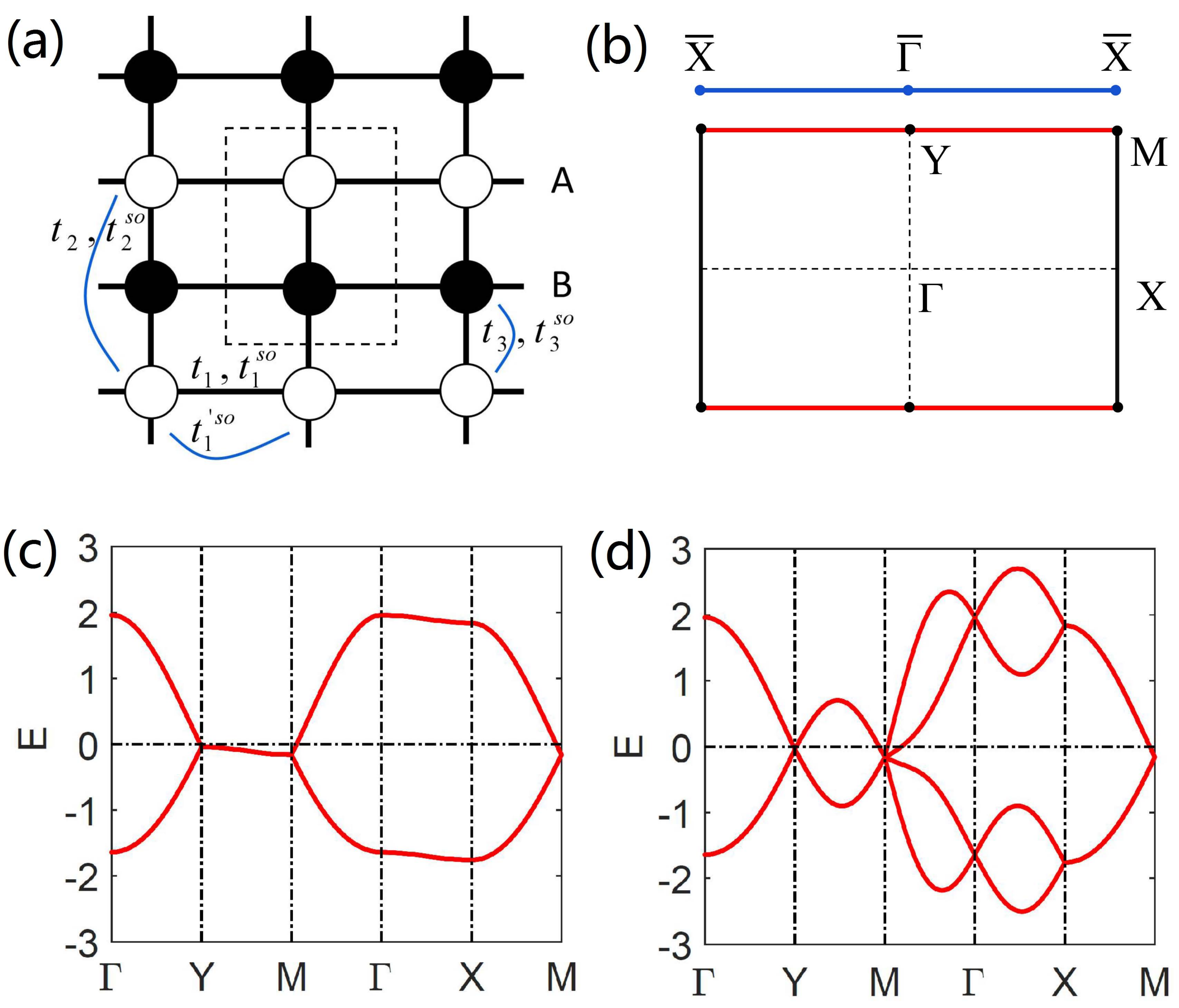}
	\caption{(a) 2D lattice with two sites A and B, which are denoted by white and black spheres, respectively. The dashed rectangular area represents the unit cell, and the blue curves mark with the hopping parameters ${t}_{i}$, ${t}^{so}_{i}$ $(i=1, 2, 3)$, and $t^{\prime so}_{1}$ represent the introduced hopping between two atoms. (b) The corresponding 2D BZ and the projected 1D BZ along the y direction, where the high-symmetry path that hosts the nodal line without SOC is marked in red, and the positions of Dirac points with SOC are denoted by black dots. Band structures without (c) and with (d) SOC.
\label{figure-2}}
\end{figure}

First, we construct a two-site tight-binding (TB) model to investigate the Dirac points in two dimensions. As shown in Fig. \ref{figure-2}(a), we consider a 2D lattice that consists of two sites A and B per unit cell. The corresponding BZ is shown in Fig. \ref{figure-2}(b). The lattice contains two non-symmorphic (glide mirror) symmetries: $\widetilde{\mathcal{M}}$$_x$=$\{\mathcal{M}_{x} | \mathbf{\tau}\}$ and $\widetilde{\mathcal{M}}$$_z$=$\{\mathcal{M}_{z} | \mathbf{\tau}\}$, where ${\mathcal{M}_x}$ and ${\mathcal{M}_z}$ are the mirror reflection of the \emph{yz} and \emph{xy} planes, respectively, and $\mathbf{\tau}=(0,\frac{1}{2},0)$ is half translation of lattice along the $\emph{y}$ axis. Note that here the $\mathcal{T}$ symmetry preserves while the $\mathcal{P}$ symmetry is broken. We assume that each site has an $\emph{s}$ orbital with two spin states, so the system has four basis sets $\ket{\emph{A},\uparrow}$, $\ket{\emph{A},\downarrow}$, $\ket{\emph{B},\uparrow}$, $\ket{\emph{B},\downarrow}$. The symmetry operators of this lattice can be represented as
\begin{equation}\label{operators}
\widetilde{\mathcal{M}}_{x} =-i\tau_{x}\sigma_{x},\quad \widetilde{\mathcal{M}}_{z} =-i\tau_{x}\sigma_{z},\quad \mathcal{T}=-i\sigma_{y}\mathcal{K},
\end{equation}
where both $\tau_{i}$ and $\sigma_{i}$ $(i=x, y, z)$ are Pauli matrices that describe the degrees of freedom for lattice and spin, respectively, and $\mathcal{K}$ is the complex conjugate operator. Constrained by above symmetries, the 4$\times$4 Hamiltonian is given by
\begin{equation}\label{Haha}
\begin{split}
&H = t_{1} {\rm cos}k_{x}\tau_{o} + t_{2} {\rm cos}k_{y}\tau_{o} +t_{3} {\rm cos}(\frac{k_y}{2})\tau_{x} + t^{so}_{1} {\rm sin}k_{x}\tau_{o}\sigma_{z}\\
&+t^{\prime so}_{1}{\rm sin}k_{x}\tau_{z}\sigma_{x} + t^{so}_{2} {\rm sin}k_{y}\tau_{z}\sigma_{y}+t^{so}_{3} {\rm cos}(\frac{k_y}{2})\tau_{y}\sigma_{y},
\end{split}
\end{equation}
where ${t}_{i}$, ${t}^{so}_{i}$ $(i=1, 2, 3)$, and $t^{\prime so}_{1}$ are hopping parameters, and more details are provided in Fig. \ref{figure-2}(a). A set of suitable values (i.e., $t_{1}=-0.06$, $t_{2}=-0.1$, $t_{3}=-1.8$, $t^{so}_{1}=-0.3$, $t^{\prime so}_{1}=-0.2$, $t^{so}_{2}=-0.4$, and $t^{so}_{3}=-0.5$) are chosen by fitting to first-principles calculations.

Based on the TB Hamiltonian in Eq. (\ref{Haha}), we calculate the band structures without and with SOC, as shown in Figs. \ref{figure-2}(c) and \ref{figure-2}(d), respectively. In the absence of SOC, the energy bands host a doubly degenerate (fourfold degenerate if the spin is considered) nodal line along the high-symmetry path Y-M. Such nodal line is protected by the $\mathcal{T}$ symmetry (with $\mathcal{T}^2=+1$) and the glide mirror line $\widetilde{\mathcal{M}}$$_x$ because their combination has the product $(\textbf{$\widetilde{\mathcal{M}}$$_x$$\mathcal{T}$})^{2}=-1$ that can give rise to the Kramers-like double degeneracy. The position of this nodal line is labeled by red in the BZ, as shown in Fig. \ref{figure-2}(b).

\begin{figure}
	\centering
	\includegraphics[scale=0.08]{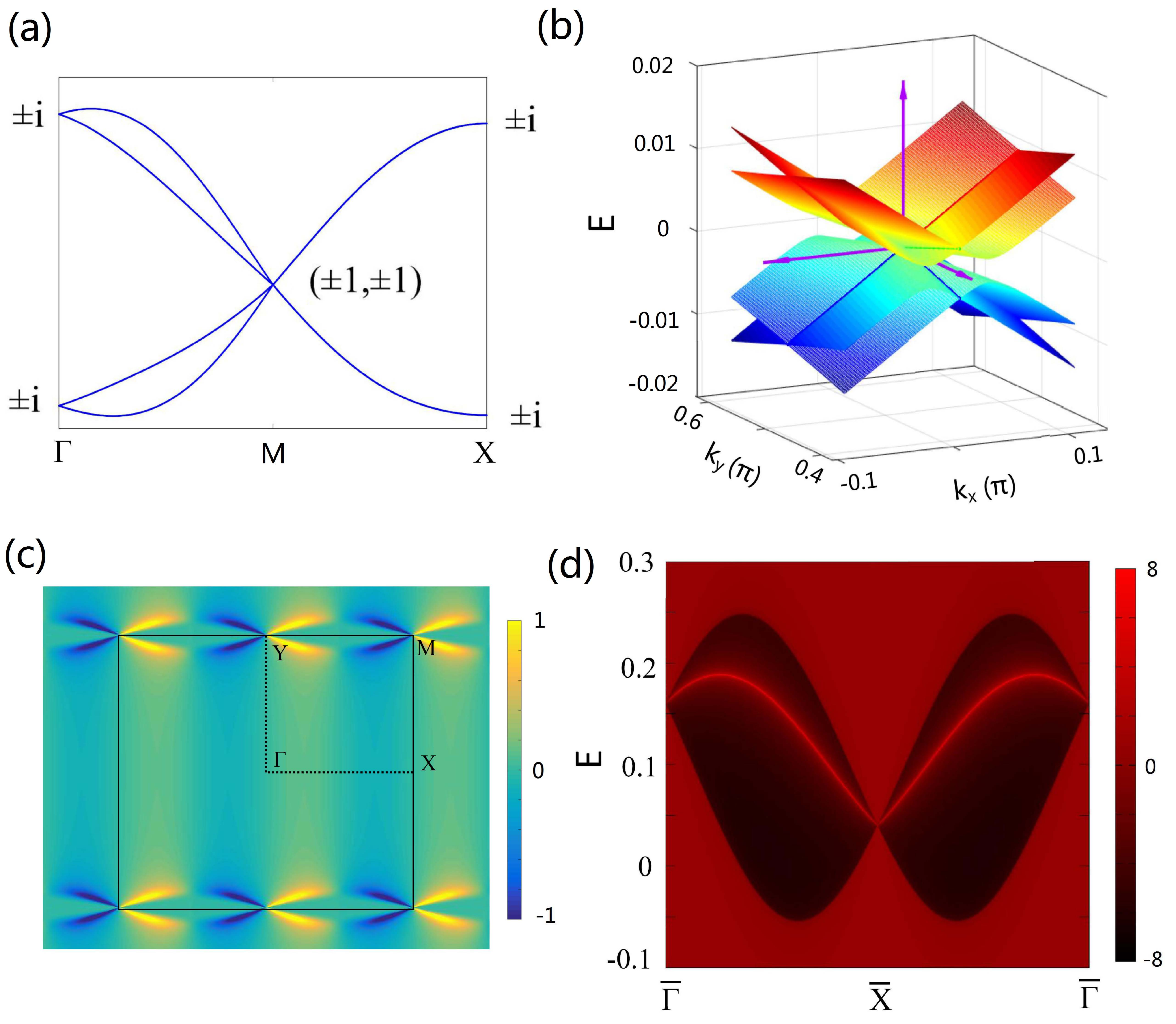}
	\caption{(a) The band structure with SOC along high-symmetry $\Gamma$-M-X lines, where $\pm$1 and $\pm$i are the eigenvalues of $\widetilde{\mathcal{M}}$$_z$. (b) 3D representation of the band structure around the M point. (c) Berry curvature distribution in the BZ. (d) The edge states in a relative scale, which are obtained on one edge of a semi-infinite ribbon along the x-direction.
\label{figure-3}}
\end{figure}

When the SOC is included, the $\mathcal{T}$ symmetry obeys $\mathcal{T}^2=-1$. Combined with the $\widetilde{\mathcal{M}}$$_x$ symmetry, we can obtain $(\textbf{$\widetilde{\mathcal{M}}$$_x$$\mathcal{T}$})^{2}=-1$ which leads to a double degeneracy along the Y-M path. Therefore, as shown in Fig. \ref{figure-2}(d), the fourfold degenerate nodal line without SOC is split into two doubly degenerate bands along the Y-M path. Since each momentum \textbf{k} is invariant under $\widetilde{\mathcal{M}}_{z}$, the Bloch states $\ket{u_\textbf{k}}$ can be chosen as the eigenstates of $\widetilde{\mathcal{M}}$$_z$$\ket{u_\textbf{k}}$=$\emph{g}$$\ket{u_\textbf{k}}$, where $\emph{g}$ have the values of $\pm$$i$$e^{i\frac{k_y}{2}}$ and are labeled at the TRIM points in Fig. \ref{figure-3}(a). According to Eq. (\ref{operators}), $\widetilde{\mathcal{M}}$$_x$ and $\widetilde{\mathcal{M}}$$_z$ anticommute with each other. Therefore, the eigenstates $\ket{u_\textbf{k}}$ and $\widetilde{\mathcal{M}}$$_x$$\ket{u_\textbf{k}}$ have the opposite $\widetilde{\mathcal{M}}$$_z$ eigenvalues $\pm$1, thus giving rise to the doubly degenerate states along the $\Gamma$-Y (or X-M) line. Next, we provide a symmetry analysis for these fourfold degenerate points. For the TRIM points Y and M, the eigenstates $\ket{u_\textbf{k}}$ and $\widetilde{\mathcal{M}}$$_x$$\ket{u_\textbf{k}}$ always accompany with their Kramers partners $\mathcal{T}$$\ket{u_\textbf{k}}$ and $\mathcal{T}$$\widetilde{\mathcal{M}}$$_x$$\ket{u_\textbf{k}}$ (see details in the Supplemental Materials (SM)\cite{SM}), guaranteeing the fourfold degenerate Dirac points (see Fig. \ref{figure-2}(d)). In short, the fourfold degenerate states are split into two doubly degenerate states along the high-symmetry lines and further degenerate into four non-degenerate states at generic points, forming the 2D Dirac points at the Y and M points. To gain an intuitive understanding of the Dirac points, we plot the 3D representation of the band structure around the M point, as shown in Fig. \ref{figure-3}(b).

To further verify these 2D Dirac points, we construct an effective four-band model at the Y point using four basis sets $\ket{u_\textbf{k}}$, $\mathcal{T}$$\widetilde{\mathcal{M}}$$_x$$\ket{u_\textbf{k}}$, $\mathcal{T}$$\ket{u_\textbf{k}}$, and $\widetilde{\mathcal{M}}$$_x$$\ket{u_\textbf{k}}$. Based on the generating operators $\widetilde{\mathcal{M}}$$_x$, $\widetilde{\mathcal{M}}$$_z$, and $\mathcal{T}$, we can obtain a generic effective Hamiltonian,
\begin{equation}\label{op111}
H(k)=g_1
\left(
  \begin{array}{cccc}
k\cdot\mu & 0\\
0 & -k\cdot\mu
  \end{array}
\right)
+T(k),
\end{equation}
where $T(k)$ =$-m_{1}k_{x}\mu_{0}\mu_{z}-m_{2}k_{y}\mu_{z}\mu_{z}$ is the distortion term. Here $g_1$, $m_1$, and $m_2$ are nonzero real constants, and $\mu_i$ are Pauli matrices (see more details in the SM\cite{SM}). Clearly, it is an exact Dirac equation modified by a distortion term $T(k)$, which can be considered as the direct sum of two Weyl equations with opposite chirality, further confirming the 2D Dirac points at Y. This effective  model is also suitable for elaborating the M point because the Y and M points share the same little group.

In our model, the $\mathcal{P}$ symmetry is broken, and the Berry curvature $\textbf{$\Omega$}_n(\textbf{k})$ is therefore locally nonzero and is given by $\textbf{$\Omega$}_n(\textbf{k})$=i $\bra{\nabla_\textbf{k}u_n(\textbf{k})}$ $\times$ $\ket{\nabla_\textbf{k}u_n(\textbf{k})}$, where $\ket{u_n(\textbf{k})}$ is the Bloch wavefunction of the \emph{n}-th band. In two dimensions, the Berry curvature has the expression
\begin{equation}\label{ch}
\textbf{$\Omega$}_n(\textbf{k}) = -\sum_{n'\ne n}2 {\rm Im} \frac{\bra{u_n(\textbf{k})} \nu_x \ket{u_{n^{'}}(\textbf{k})}  \bra{u_{n}^{'}(\textbf{k})} \nu_y \ket{u_n(\textbf{k})}}{(\varepsilon_{n'\textbf{k}}-\varepsilon_{n\textbf{k}})^2}
\end{equation}
where $\nu_\alpha$=$\frac{1}{\hbar}$ $\frac{\partial H}{\partial k_\alpha}$($\alpha = x, y$) are velocity matrices and $\varepsilon_{n\textbf{k}}$ are the eigenvalues of the \emph{n}-th band. Based on Eq. (\ref{ch}), the Berry curvature for the TB model is calculated, and it exhibits a clover-type pattern, as shown in Fig. \ref{figure-3}(c). The $\mathcal{T}$ symmetry gives
\begin{equation}\label{cher1}
\Omega_n(k_x,k_y) =-\Omega_n(-k_x,-k_y).
\end{equation}
Under the $\widetilde{\mathcal{M}}$$_x$ symmetry, we can obtain
\begin{equation}\label{cher2}
\Omega_n(k_x,k_y) =-\Omega_n(-k_x,k_y).
\end{equation}
The combination operator $\widetilde{\mathcal{M}}$$_x$$\mathcal{T}$ requires
\begin{equation}\label{cher3}
\Omega_n(k_x,k_y) =\Omega_n(k_x,-k_y).
\end{equation}
Clearly, under the constraint of the $\mathcal{T}$ symmetry, the Berry curvature is an odd function of $k_x$ and $k_y$. Meanwhile, the Berry curvature is odd with respect to $k_x$  due to the $\widetilde{\mathcal{M}}$$_x$ symmetry and is even along $k_y$ constrained by the $\widetilde{\mathcal{M}}$$_x$$\mathcal{T}$ symmetry. The pattern of the calculated Berry curvature is in excellent  agreement with the symmetry constraints, as shown in Fig. \ref{figure-3}(c). It exhibits nonzero Berry curvature distribution with large anisotropy in the vicinity of Dirac points. The Berry curvature is well separated in the BZ, which may favor large nonlinear Hall effect\cite{nonlinerhalleffect,nonhallexp00}. Besides, the inequivalent valleys at the Y and M points with fourfold degeneracy offer a unique platform to investigate valleytronics in topological semimetals beyond graphene.

Such nonzero local Berry curvature can lead to nontrivial edge states. Employing the iterative Green's function method\cite{greensfunction,WU2017}, we calculate the local density of states (LDOS) on one edge of a semi-infinite nanoribbon along the x direction. The edge states in a relative scale are shown in Fig. \ref{figure-3}(d). It can be clearly seen that two visible edge states spanning the whole 1D BZ connect one pair of projected Dirac points, characterizing the 2D Dirac points without $\mathcal{P}$ symmetry. This is completely different from Weyl points and $\mathcal{PT}$ symmetry-involved Dirac points. It is worth noting that the edge states disappear along the y direction because the opposite Berry curvatures contributed from $k_x$ and $-k_x$ cancel each other out when it projects along the y direction. Such direction-dependent edge states have potential applications in anisotropic electronic devices.

\begin{figure}
	\centering
	\includegraphics[scale=0.092]{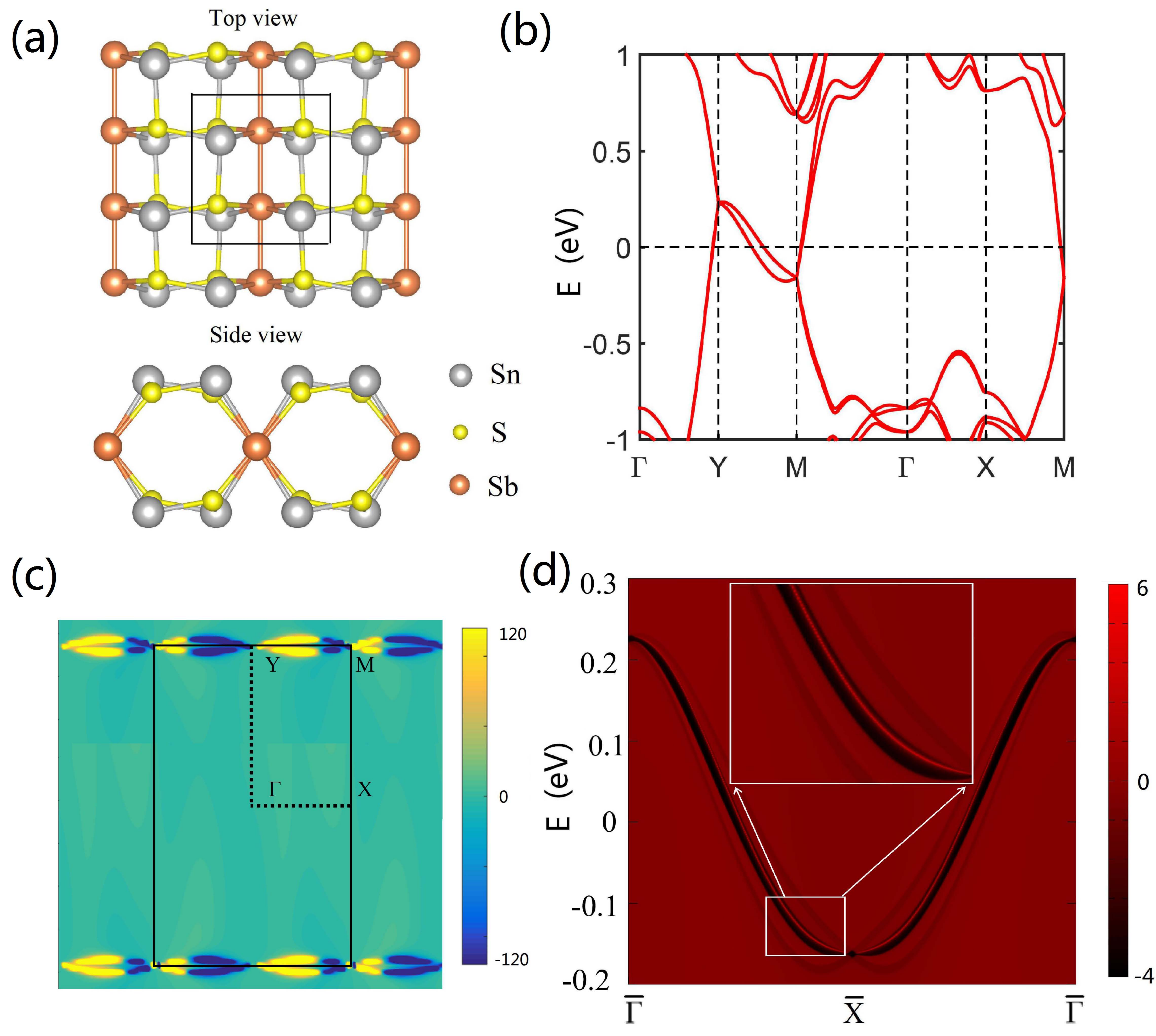}
	\caption{(a) Top (upper panel) and side (lower panel) views of monolayer SbSSn, and the unit cell is marked by the black rectangle. (b) Calculated band structure of SbSSn with SOC. (c) Berry curvature distribution in the BZ. (d) The edge states in a relative scale, which are obtained from one edge of a semi-infinite ribbon along the x direction.
\label{figure-4}}
\end{figure}

In addition to the above discussions based on the TB model, we identify a family of candidates as ideal 2D Dirac semimetals using first-principles calculations. Here, we only focus on the monolayer SbSSn, and other candidates are summarized in the SM\cite{SM}. The monolayer SbSSn contains three atomic layers, where two slightly puckered S and Sn layers are well separated by the middle Sb atomic layer, as shown in Fig. \ref{figure-4}(a). The optimized lattice parameters and atomic positions are given in the SM\cite{SM}. To examine the dynamical stability, we calculate the phonon spectra (see the SM\cite{SM}) of SbSSn. There is no imaginary frequency found in the BZ, implying its dynamical stability.

The electronic band structure of SbSSn with SOC is plotted in Fig. \ref{figure-4}(b). Clearly, there are two inequivalent fourfold degenerate crossing points located at Y and M, forming electron and hole valleys around the Fermi level. The little groups of the crossing points at the Y and M points have only one four-dimensional (4D) irreducible representation\cite{Bilbaoserve}, guaranteeing the fourfold degenerate 2D Dirac points. Moving along the high symmetry lines, the 4D irreducible representation degenerates into two dimensions and is further reduced to one dimension at generic points, which shows excellent consistency with the above symmetry arguments. The Berry curvature distribution and LDOS projected on one edge of the semi-infinite ribbon along the x direction are shown in Figs. \ref{figure-4}(c) and \ref{figure-4}(d), respectively. As expected, the Berry curvature pattern obeys the symmetry conditions, and two edge states connect one pair of Dirac points spanning over the whole BZ.


In conclusion, using symmetry arguments, we present that 2D Dirac semimetals can exist without $\mathcal{P}$ symmetry. These Dirac points are protected by $\mathcal{T}$ symmetry, glide mirror plane $\widetilde{\mathcal{M}}$$_z$, and glide mirror line $\widetilde{\mathcal{M}}$$_x$. Different from $\mathcal{PT}$ symmetry-involved 2D Dirac points, our proposed model without $\mathcal{P}$ symmetry features nonzero Berry curvature in the vicinity of the Dirac points. It also hosts two edge states terminated at the projections of two Dirac points spanning the whole 1D BZ. Furthermore, using high-throughput first-principles calculations, we find a family of 2D Dirac candidates, which has nonzero Berry curvature near the crossing points and visible edge states, making their exotic transport properties easy to be observed in experiments. Our findings show that 2D Dirac semimetals without $\mathcal{P}$ symmetry exhibit a wide range of physical properties different from graphene, e.g., non-linear Hall effect, and have potential applications in low-dimensional quantum transport devices.

~~~\\
~~~\\
 This work is supported by the National Natural Science Foundation of China (NSFC, Grant Nos. 11974160, 11674148 and 11334004), the Guangdong Natural Science Funds for Distinguished Young Scholars (No. 2017B030306008), and the Center for Computational Science and Engineering at Southern University of Science and Technology.\\
~~~\\

Y.J.J and B.B.Z equally contributed to this work.


\end{document}